\documentclass[fleqn,10pt]{wlscirep}
\usepackage{color}
\usepackage{amssymb,amsmath,amsthm}

\title{The complementarity relations of quantum coherence in quantum information processing}

\author{Fei Pan}
\author[1,*]{Liang Qiu}
\author{Zhi Liu}
\affil{School of Physics, China University of Mining and Technology, Xuzhou 221116, China}

\affil[*]{lqiu@cumt.edu.cn}

\begin{abstract}
We establish two complementarity relations for the relative entropy of coherence in quantum information processing, i.e., quantum dense coding and teleportation. We first give an uncertainty-like expression relating local quantum coherence to the capacity of optimal dense coding for bipartite system. The relation can also be applied to the case of dense coding by using unital memoryless noisy quantum channels. Further, the relation between local quantum coherence and teleportation fidelity for two-qubit system is given.
\end{abstract}
\begin{document}

\flushbottom
\maketitle
% * <john.hammersley@gmail.com> 2015-02-09T12:07:31.197Z:
%
%  Click the title above to edit the author information and abstract
%
\thispagestyle{empty}

%\noindent Please note: Abbreviations should be introduced at the first mention in the main text - no abbreviations lists. Suggested structure of main text (not enforced) is provided below.

\section*{Introduction}
Quantum coherence, which arises from quantum superposition, is a fundamental feature of quantum mechanics, and it is also an essential ingredient in quantum information and computation \cite{Nielsen}. Furthermore, in some emergent fields, such as quantum metrology \cite{Demkowicz,Giovannetti1}, nanoscale thermodynamics \cite{Aberg,Narasimhachar,Cwiklinski,Lostaglio,Lostaglio1} and quantum biology \cite{Plenio,Lloyd,Li,Huelga}, quantum coherence plays a central role.

The information-theoretic quantification of quantum coherence is a successful application of quantum resource theory \cite{Baumgratz}. Baumgratz et al. proposed the basic notions of incoherent states, incoherent operations and a series of necessary conditions any measures of coherence should satisfy. In this sense, coherence is defined as the resource relative to the set of incoherent operations. According to the postulates in the framework, relative entropy of coherence \cite{Baumgratz}, $l_{1}$-norm of coherence \cite{Baumgratz} and other coherence metrics \cite{Streltsov,Winter,Chitambar,Yuan,Du} have been put forward. Based on coherence measures, the relations between quantum coherence and other resources\cite{Streltsov,Yao,Xi}, the complementarity relations of quantum coherence \cite{Cheng} and other properties of quantum coherence \cite{Streltsov3,deVicente} have been investigated.
Mainly due to the interest aroused by the resource theory of quantum coherence, there are several attempts at understanding the role of coherence as a resource for quantum protocols. For example, in the incoherent quantum state merging, which is the same as standard quantum state merging up to the fact that one of the parties has free access to local incoherent operations only and has to consume a coherent resource for more general operations, the entanglement-coherence sum is non-negative, and no merging procedure can gain entanglement and coherence at the same time \cite{Streltsov1}. Perfect incoherent teleportation of an unknown state of one qubit is possible with one singlet and two bits of classical communications \cite{Streltsov2}. Here, the incoherent teleportation is the same as standard teleportation up to the fact that local operations and classical communications are replaced by local incoherent operations and classical communications. Furthermore, the notion of coherence as a symmetry relative to a group of translations naturally shows up in the context of quantum speed limits because the speed of evolution is itself a measure of asymmetry relative to time translations \cite{Marvian}.

As we know, both quantum coherence and entanglement closely relate to quantum superposition. Moreover, many quantum information protocols, such as dense coding~\cite{Bennett} and teleportation~\cite{Bennett1}, would be impossible without the assistance of entanglement. Therefore, inspired by work on entanglement, we want to directly relate quantum coherence with the protocols of quantum information. Specifically, we want to give the quantitative relation between quantum coherence and the dense coding capacity or teleportation fidelity.

In a realistic scenario, the inevitable interactions between the system and the environment always lead to decoherence of the system and the rapid destruction of quantum properties. The dynamics of quantum coherence has been extensively investigated~\cite{Addis,Bromley,Yu,Bhattacharya}. Dense coding in the presence of noise has also attracted much attention~\cite{Shadman,Shadman1,Shadman2,Das,Das1,Wangxin,Liu}, as well as teleportation~\cite{Bowen,Albeverio,Oh,Taketani,Knoll,Fortes,Fortes1}. In particular, dense coding for the case that the subsystems of the entangled resource state have to pass a noisy unital quantum channel between the sender and the receiver is considered in Ref.~\citen{Shadman}. We try to apply the quantitative relation between quantum coherence and the dense coding capacity to this special case. Moreover, we will explore whether the quantitative relations between quantum coherence and the dense coding capacity, and that between quantum coherence and teleportation fidelity can be generalized to the general noisy maps.

In the present work, we will establish a complementarity relation between quantum coherence and the optimal dense coding capacity, and also relate quantum coherence to teleportation fidelity in the form of a complementarity relation. Here, quantum coherence is measured by the relative entropy of coherence.

\section*{Results}
\subsection*{Relating quantum coherence to optimal dense coding and teleportation}
In this section, we will investigate the relation between quantum coherence and the optimal dense coding, and that between quantum coherence and teleportation.

The definition of relative entropy of coherence $C_{\rm re}$ \cite{Baumgratz} is
\begin{equation}
C_{\rm re}(\rho)=\min_{\delta\in\mathcal{I}}S(\rho\|\delta),
\end{equation}
where $S(\rho\|\delta)={\rm tr}\rho(\log_{2}\rho-\log_{2}\delta)$ is the relative entropy, $\mathcal{I}$ is the set of all incoherent states and all density operators $\delta\in\mathcal{I}$ are of the form \cite{Baumgratz}
\begin{equation}
\delta=\sum\limits_{i=1}^{d}\delta_{i}|i\rangle\langle i|,
\end{equation}
with $\{|i\rangle\}_{i=1,\ldots,d}$ being a particular basis of the $d$-dimensional Hilbert space $\mathcal{H}$. In the definition of relative entropy of coherence, the minimum is attained if and only if $\delta=\rho^{diag}$ with $\rho^{diag}$ being the diagonal part of $\rho$. $C_{\rm re}$ satisfies the four postulates given in Ref.~\citen{Baumgratz} which are the conditions that a measure of quantum coherence should satisfy. Based on the definition, we can establish the complementarity relation between local quantum coherence and the optimal dense coding.

\subsubsection*{Relating quantum coherence to optimal dense coding}
For a bipartite quantum state $\rho_{AB}$ on two $d$-dimensional Hilbert spaces $\mathcal{H}_{A}^{d}\otimes\mathcal{H}_{B}^{d}$ with $\rho_{B}={\rm tr}_{A}(\rho_{AB})$ being the reduced density matrix of the subsystem $B$, we have the following theorem.

\textbf{Theorem 1} The sum of the optimal dense coding capacity of the state $\rho_{AB}$ and quantum coherence of the reduced state $\rho_{B}$ is always smaller than $2\log_{2}d$, i.e.,
\begin{equation}
\chi(\rho_{AB})+C_{\rm re}(\rho_{B})\leq2\log_{2}d,\label{eq3}
\end{equation}
where $\chi(\rho_{AB})$ is the optimal dense coding capacity of the state $\rho_{AB}$.

\begin{proof} The $d^2$ signal states generated by mutually orthogonal unitary transformations with equal probabilities will yield the maximal $\chi$ \cite{Hiroshima,Qiu2017}. The mutual orthogonal unitary transformations are given as
\begin{equation}
U_{m,n}|j\rangle={\rm exp}\left(i\frac{2\pi}{d}mj\right)|j+n({\rm mod}\ d)\rangle,\label{eq4}
\end{equation}
where integers $m$ and $n$ range from $0$ to $d-1$. The ensembles generated by the unitary transformations with equal probabilities $p_{m,n}$ can be denoted as $\varepsilon^{\ast}=\{(U_{m,n}^{A}\otimes I_{d}^{B})\rho_{AB}(U_{m,n}^{A\dag}\otimes I_{d}^{B});\ p_{m,n}=1/d^2\}_{m,n=0}^{d-1}$. The average state of the ensembles is
\begin{equation}
\overline{\rho^{\ast}_{AB}}=\frac{1}{d^2}\sum\limits_{m,n}^{d-1}(U_{m,n}^{A}\otimes I_{d}^{B})\rho_{AB}(U_{m,n}^{A\dag}\otimes I_{d}^{B}).
\end{equation}
Here, $I_{d}^{B}$ is the $d$-dimensional identity matrix in the subsystem $B$. Accordingly, the capacity of the optimal dense coding can be given as \cite{Hiroshima}
\begin{equation}
\chi(\rho_{AB})=S\left(\overline{\rho^{\ast}_{AB}}\right)-S\left(\rho_{AB}\right).
\end{equation}
Based on the result in Ref.~\citen{Hiroshima}, i.e., $\overline{\rho^{\ast}_{AB}}=I_{d}^{A}\otimes\rho_{B}/d$, we have
\begin{equation}
S\left(\overline{\rho^{\ast}_{AB}}\right)=-{\rm tr}\left(\overline{\rho^{\ast}_{AB}}\log_{2}\overline{\rho^{\ast}_{AB}}\right)=-{\rm tr}(I_{d}){\rm tr}\left(\frac{\rho_{B}}{d}\log_{2}\frac{\rho_{B}}{d}\right)=S(\rho_{B})+\log_{2}d.
\end{equation}

For the reduced state $\rho_{B}$ of the subsystem $B$, $C_{\rm re}(\rho_{B})=S(\rho_{B}^{\rm diag})-S(\rho_{B})$, and $S(\rho_{B}^{\rm diag})\leq\log_{2}d$. Therefore, $C_{\rm re}(\rho_{B})\leq\log_{2}d-S(\rho_{B})$, from which we have
\begin{equation}
C_{\rm re}(\rho_{B})+S(\rho_{B})\leq\log_{2}d.
\end{equation}
Now, we consider the sum of the optimal dense coding capacity of the whole system $AB$ and quantum coherence of the subsystem $B$
\begin{equation}
\chi(\rho_{AB})+C_{\rm re}(\rho_{B})=S(\rho_{B})+\log_{2}d-S(\rho_{AB})+C_{\rm re}(\rho_{B})\leq\log_{2}d+\log_{2}d-S(\rho_{AB})\leq2\log_{2}d,
\end{equation}
where the first inequality is attained because of the fact given in Eq. (8), and the second inequality is obtained due to $S(\rho_{AB})\geq0$. This completes the proof. \end{proof}

For the particular case that the shared entangled state is the Bell state, $\chi(\rho_{AB})=2$ and $C_{\rm re}(\rho_{B})=0$, and the sum of them equals to $2$, which just equals to the right hand side of Eq.~(\ref{eq3}).

The inequality given in Eq.~(\ref{eq3}) indicates that the greater local quantum coherence is, the smaller capacity of the optimal dense coding will be. In other words, if the system $AB$ is used to perform dense coding as much as possible, quantum coherence of the subsystem $B$ would pay for the dense coding capacity of the whole system. The physical reason is that dense coding is based on entanglement, and would be impossible without the assistance of entangled states. The results given in Ref. \citen{Xi} show that entanglement of the whole system and quantum coherence of a subsystem are complementary to each other. That is, an increase in one leads to a decrease in the other. For example, for a Bell state, an incoherent state of the subsystem $B$ will be acquired if qubit $A$ is traced over. On the contrary, creating a superposition on a subsystem to have maximum coherence on it will exclude entanglement between subsystems.

In Ref.~\citen{Streltsov2}, the task of incoherent quantum state merging is introduced and the amount of resources needed for it is quantified by an entanglement-coherence pair. It is found that the entanglement-coherence sum is non-negative, in other words, no merging procedure can gain entanglement and coherence at the same time. From the results given in this paper, the sum of the optimal dense coding capacity and quantum coherence is upper bounded by a definite value, i.e., there is a trade-off between the dense coding capacity and quantum coherence. It should be noted that dense coding is based on entanglement, and the former would be impossible when the latter is absent. In this sense, the result given in Eq.~(\ref{eq3}) is consistent with those presented in Ref.~\citen{Streltsov2}.

The result given in Theorem 1 can also be extended to the case of dense coding by using unital memoryless noise quantum channels. The unital noisy channels acting on Alice's and Bob's systems are described by the completely positive map $\Lambda(\rho)=\sum_{i}K_{i}\rho K_{i}^{\dag}$, where $\sum_{i}K_{i}^{\dag}K_{i}=I$ corresponds to trace preservation, and $\sum_{i}K_{i}K_{i}^{\dag}=I$ guarantees the unital property, i.e., $\Lambda(I)=I$. Here, $K_{i}$ denotes the Kraus operators. In Ref.~\citen{Shadman}, the authors found that the encoding with the equally probable operators $U_{m,n}$, as given in Eq.~(\ref{eq4}), is optimal for the states of which the von Neumann entropy after the channel action is independent of unitary encoding. In other words, the states satisfy
\begin{equation}
S(\Lambda_{AB}(\rho))=\frac{1}{d^2}\sum_{m,n=0}^{d-1}S(\Lambda_{AB}(\rho_{m,n})),\label{eq10}
\end{equation}
where $\rho_{m,n}=(U_{m,n}^{A}\otimes I_{d}^{B})\rho(U_{m,n}^{A\dag}\otimes I_{d}^{B})$. The corresponding dense coding capacity can also be given by $\chi(\Lambda_{AB}(\rho_{AB}))=S(\overline{\rho_{AB}})-S(\Lambda_{AB}(\rho_{AB}))$, where $\overline{\rho_{AB}}$ is the average of the ensemble after encoding with the equally probable unitaries $U_{m,n}$ and after the channel action. That is, $\overline{\rho_{AB}}$ is the average state of the ensemble $\{\Lambda_{AB}[(U_{m,n}^{A}\otimes I_{d}^{B})\rho_{AB}(U_{m,n}^{A\dag}\otimes I_{d}^{B})];\ p_{i}=\frac{1}{d^2},\}_{m,n=0}^{d-1}$. Based on the fact that $\overline{\rho_{AB}}=I_{A}\otimes\Lambda_{B}(\rho_{B}/d)$~\cite{Shadman}, $\chi(\Lambda_{AB}(\rho_{AB}))=\log_{2}d+S(\Lambda_{B}(\rho_{B}))-S(\Lambda_{AB}(\rho_{AB}))$. Following the proof process of Theorem 1, one can easily obtain $\chi(\Lambda_{AB}(\rho_{AB}))+C_{\rm re}(\Lambda_{B}(\rho_{B}))\leq2\log_{2}d$, which indicates our result in Eq.~(\ref{eq3}) applying to the case of dense coding by using unital memoryless noise quantum channels.

Now, we consider an example of two-sided depolarizing channel~\cite{Shadman}. Alice firstly prepares the bipartite state $\rho_{AB}$, and sends one part of it, i.e., $B$, via a noisy channel $\Lambda_{B}$ to the receiver, Bob, so as to establish the shared state for dense coding. Subsequently, Alice does the local unital encoding and then sends her part of the state, i.e., $A$, via the noisy channel $\Lambda_{A}$ to Bob. The two-sided $d$-dimensional depolarizing channel is defined as
\begin{equation}
\Lambda_{AB}^{\rm dep}(\rho_{AB})=\sum\limits_{\mu,\nu,\widetilde{\mu},\widetilde{\nu}=0}^{d-1}q_{\mu\nu}q_{\widetilde{\mu}\widetilde{\nu}}(V_{\mu\nu}\otimes V_{\widetilde{\mu}\widetilde{\nu}})\rho_{AB}(V_{\mu\nu}^{\dag}\otimes V_{\widetilde{\mu}\widetilde{\nu}}^{\dag}),\label{eq11}
\end{equation}
with the probability parameters $q_{\mu\nu}=1-(d^2-1)p/d^2$ for $\mu=\nu=0$, otherwise $q_{\mu\nu}=p/d^2$. The operators $V_{\mu\nu}$ read
\begin{equation}
V_{\mu\nu}=\sum\limits_{k=0}^{d-1}\exp{\left(\frac{2i\pi k\nu}{d}\right)}|k\rangle\langle k+\mu({\rm mod}\ d)|.
\end{equation}
It is proved that the von Neumann entropy of a state, which is sent through the two-sided depolarizing channels, is independent of any local unitary transformations that were performed before the action of the channel, i.e., the condition given in Eq.~(\ref{eq10}) is satisfied~\cite{Shadman}.

\begin{figure}[h]%[tbp]
\begin{center}
\includegraphics[scale=1.0]{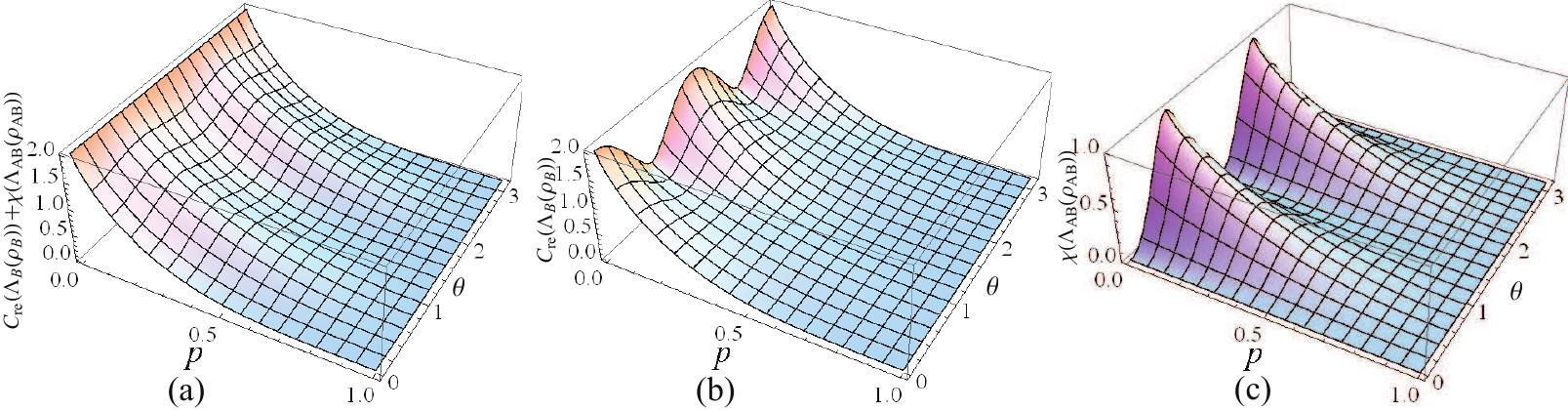}
\end{center}
\caption{(a) The sum of the relative entropy of coherence for subsystem $B$ $C_{\rm re}(\Lambda_{B}(\rho_{B}))$ and the dense coding capacity $\chi(\Lambda_{AB}(\rho_{AB}))$, (b) $C_{\rm re}(\Lambda_{B}(\rho_{B}))$, and (c) $\chi(\Lambda_{AB}(\rho_{AB}))$ as functions of the state parameter $\theta$ and the noise parameter $p$.}\label{fig1}
\end{figure}

\begin{figure}[h]%[tbp]
\begin{center}
\includegraphics[scale=1.0]{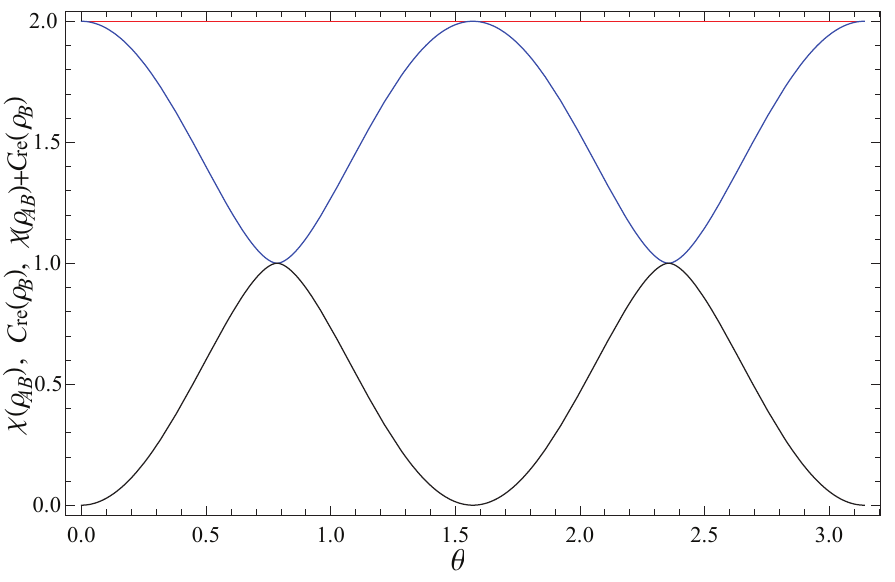}
\end{center}
\caption{The sum of the relative entropy of coherence for subsystem $B$ $C_{\rm re}(\rho_{B})$ and the dense coding capacity $\chi(\rho_{AB})$ (Red line), $C_{\rm re}(\rho_{B})$ (Black line), and $\chi(\rho_{AB})$ (Blue line) versus the state parameter $\theta$ for a fixed value of $p=0$.}\label{fig2}
\end{figure}

Specific to the case that Alice and Bob have the two-sided $2$-dimensional depolarizing channel for the transfer of the qubit states, the initial resource state is chosen as $|\phi\rangle_{AB}=\cos{\theta}|\Phi^{+}\rangle_{AB}+\sin{\theta}|\Psi^{+}\rangle_{AB}$, where $\theta\in(0,\pi)$, and $|\Phi^{+}\rangle=\frac{1}{\sqrt{2}}(|00\rangle+|11\rangle)$, $|\Psi^{+}\rangle=\frac{1}{\sqrt{2}}(|01\rangle+|10\rangle)$ are the Bell states. After sending the qubit $B$ to Bob via the depolarizing channel, Alice implements the local unital encoding and then sends the qubit $A$ to Bob via the depolarizing channel too. The dense coding capacity $\chi(\Lambda_{AB}(\rho_{AB}))$ and the relative entropy of coherence $C_{\rm re}(\Lambda_{B}(\rho_{B}))$ can be straightforwardly calculated, however, the expressions of them are analytically messy, and thus we have chosen to simply plot the exactly numerical results. In Fig.~\ref{fig1}, we plot the evolutions of $\chi(\Lambda_{AB}(\rho_{AB}))+C_{\rm re}(\Lambda_{B}(\rho_{B}))$, $\chi(\Lambda_{AB}(\rho_{AB}))$ and $C_{\rm re}(\Lambda_{B}(\rho_{B}))$ as functions of the state parameter $\theta$ and the noise parameter $p$. From Fig.~\ref{fig1}(a), it is found that $\chi(\Lambda_{AB}(\rho_{AB}))+C_{\rm re}(\Lambda_{B}(\rho_{B}))\leq2$ is always satisfied, which indicates the result given in Theorem $1$ is validated. This can be appreciated in Fig.~\ref{fig1}(b) and~\ref{fig1}(c), where $\chi(\Lambda_{AB}(\rho_{AB}))$ reaches its maximum value while $C_{\rm re}(\Lambda_{B}(\rho_{B}))$ gets its minimum value, or vice versa. The underlying physical mechanism is that the dense coding capacity is much greater when the two-qubit state is much more entangled, while the coherence of the subsystem is much smaller. This physical explanation is verified in Fig.~\ref{fig2}, where we plot $\chi(\Lambda_{AB}(\rho_{AB}))+C_{\rm re}(\Lambda_{B}(\rho_{B}))$, $\chi(\Lambda_{AB}(\rho_{AB}))$ and $C_{\rm re}(\Lambda_{B}(\rho_{B}))$ versus $\theta$ for $p=0$. For the particular cases of $\theta=\pi/4$ and $3\pi/4$, $|\phi\rangle_{AB}=\frac{1}{\sqrt{2}}(|0\rangle+|1\rangle)_{A}\otimes\frac{1}{\sqrt{2}}(|0\rangle+|1\rangle)_{B}$ and $\frac{1}{\sqrt{2}}(|0\rangle-|1\rangle)_{A}\otimes\frac{1}{\sqrt{2}}(|0\rangle-|1\rangle)_{B}$, respectively. The subsystem $B$ has the maximum value of coherence $C_{\rm re}(\rho_{B})=1$ when the two-qubit state is the product state and is useless for dense coding. On the contrary, for the cases of $\theta=0$ and $\pi/2$, $|\phi\rangle_{AB}=|\Phi\rangle_{AB}$ and $|\Psi\rangle_{AB}$, respectively, and the dense coding capacity gets its maximum value $\chi(\rho_{AB})=2$ for both of them. At these points, the two-qubit states are maximally entangled, and the subsystem has no coherence.

The relation between quantum coherence and dense coding has been given in Eq.~(\ref{eq3}), and in the following, we will relate quantum coherence to teleportation.

\subsubsection*{Relating quantum coherence to teleportation}
For an arbitrary two-qubit mixed state $\rho_{AB}$ with $\rho_{A}={\rm tr}_{B}(\rho_{AB})$ being the reduced state of the subsystem $A$, we have the following theorem.

\textbf{Theorem 2} For any two-qubit state
\begin{equation}
h\left(\frac{1+\sqrt{1-[3F(\rho_{AB})-2]^2}}{2}\right)+C_{\rm re}(\rho_{A})\leq1,\label{eq13}
\end{equation}
where $h(x)=-x\log_{2}x-(1-x)\log_{2}(1-x)$ is the binary entropy, $F(\rho_{AB})$ is the teleportation fidelity of the state $\rho_{AB}$ and $C_{\rm re}(\rho_{A})$ denotes quantum coherence of the subsystem $A$. Here, we just consider the case where the state $\rho_{AB}$ is useful for teleportation, which means $F(\rho_{AB})\geq2/3$.

\begin{proof} In the proof, the subscripts are omitted in the case that it does not cause confusion. For a two-qubit state, the relation between the teleportation fidelity $F(\rho)$ and negativity $N(\rho)$ is $3F(\rho)-2\leq N(\rho)$ \cite{Verstraete}, while negativity is related to concurrence $C(\rho)$ as $N(\rho)\leq C(\rho)$ \cite{Verstraete1}. Combining the two relations, one can obtain $3F(\rho)-2\leq N(\rho)\leq C(\rho)$. $F(\rho)\geq2/3$ leads to all of them being larger than $0$, so the square of them also obey the rules, i.e., $[3F(\rho)-2]^2\leq N^{2}(\rho) \leq C^{2}(\rho)$. Subsequently, the following expression exists
\begin{equation}
\frac{1+\sqrt{1-[3F(\rho)-2]^2}}{2}\geq\frac{1+\sqrt{1-N^{2}(\rho)}}{2}\geq\frac{1+\sqrt{1-C^{2}(\rho)}}{2}\geq\frac{1}{2}.
\end{equation}
The last inequality can be acquired based on the fact that concurrence $C(\rho)$ for two-qubit state runs from $0$ to $1$.

As known to all, $h(x)$ is a monotonically decreasing function in the interval $[1/2,1]$, thus one can obtain
\begin{equation}
h\left(\frac{1+\sqrt{1-[3F(\rho)-2]^2}}{2}\right)\leq h\left(\frac{1+\sqrt{1-N^{2}(\rho)}}{2}\right)\leq h\left(\frac{1+\sqrt{1-C^{2}(\rho)}}{2}\right)=E_{F}(\rho),\label{eq15}
\end{equation}
where $E_{F}(\rho)$ is the entanglement of formation of the state $\rho_{AB}$.

For any bipartite state $\rho_{AB}$, entanglement of formation and quantum coherence obey the relation \cite{Xi}
\begin{equation}
E_{F}(\rho_{AB})+C_{\rm re}(\rho_{A})\leq\log_{2}d_{A}.\label{eq16}
\end{equation}

Combining Eqs.~(\ref{eq15}) with ~(\ref{eq16}), and specializing to the two-qubit state, i.e., $d_{A}=2$, it is easy to complete the proof.\end{proof}

The inequality given in Eq.~(\ref{eq13}) indicates that the greater the teleportation fidelity is, the smaller local quantum coherence will be. That is to say, quantum coherence of the subsystem should pay for teleportation fidelity of the whole system. The reason for this result is that teleportation relies on entanglement. However, quantum coherence of the subsystem and entanglement of the whole system are complementary to each other.

For the particular case that the Bell state is utilized to perform teleportation, $F(\rho_{AB})=1$ leads to $h\left(\frac{1+\sqrt{1-[3F(\rho_{AB})-2]^2}}{2}\right)=1$ while $C_{\rm re}(\rho_{A})=0$. Thus, $h\left(\frac{1+\sqrt{1-[3F(\rho_{AB})-2]^2}}{2}\right)+C_{\rm re}(\rho_{A})$ equals to $1$.

\begin{figure}[h]%[tbp]
\begin{center}
\includegraphics[scale=1.0]{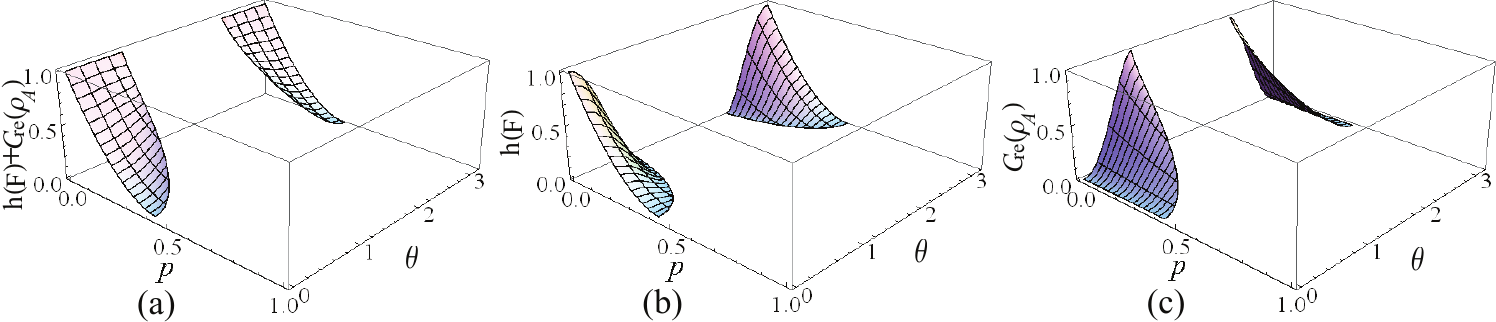}
\end{center}
\caption{(a) The sum of $h(F)$ and the relative entropy of coherence for the subsystem $A$ $C_{\rm re}(\rho_{A})$, (b) $h(F)$, and (c) $C_{\rm re}(\rho_{A})$ as functions of the state parameter $\theta$ and the noise parameter $p$. In the plot, we only consider the case of $F>2/3$.}\label{fig3}
\end{figure}

\begin{figure}[h]%[tbp]
\begin{center}
\includegraphics[scale=1.0]{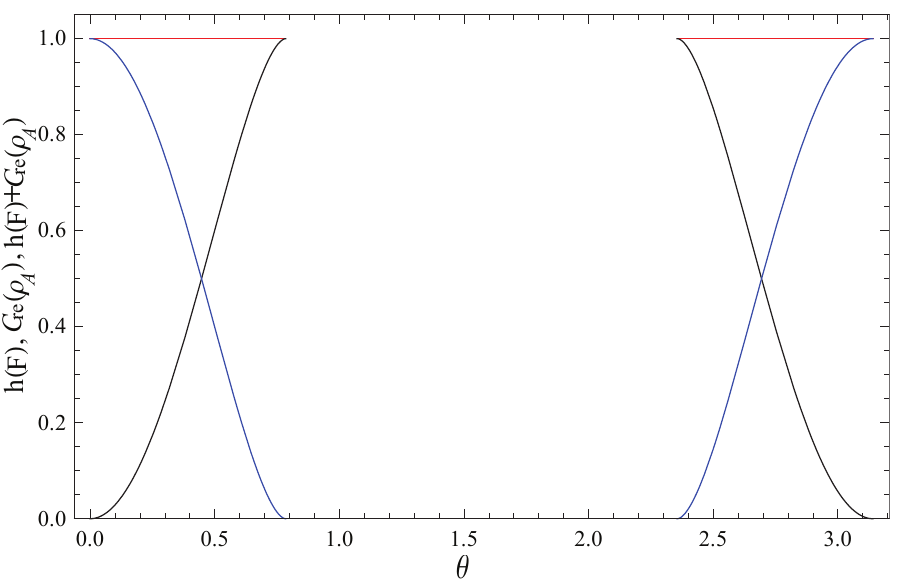}
\end{center}
\caption{The sum of $h(F)$ and the relative entropy of coherence for the subsystem $A$ $C_{\rm re}(\rho_{A})$ (Red line), $h(F)$ (Blue line), and $C_{\rm re}(\rho_{A})$ (Black line) versus the state parameter $\theta$ for a fixed value of $p=0$. In the plot, we only consider the case of $F>2/3$.}\label{fig4}
\end{figure}

Now, we investigate the example of two-qubit state $|\phi\rangle_{AB}=\cos{\theta}|\Phi\rangle_{AB}+\sin{\theta}|\Psi\rangle_{AB}$ with $\theta\in(0,\pi)$, which is distributed to Alice and Bob through the $2$-dimensional depolarizing channels. According to the Eq.~(\ref{eq11}), one can obtain the output state $\Lambda_{AB}(\rho_{AB})$, which will be considered as the resource state for implementing teleportation. The unknown state of qubit $a$ to be teleported is assumed to be $|\psi\rangle_{a}=\cos{(\alpha/2)}\exp{(i\beta/2)}|0\rangle+\sin{(\alpha/2)}\exp{(-i\beta/2)}|1\rangle$, where $\alpha\in(0,\pi),\ \beta\in(0,2\pi)$. Bob can get the teleported state $\rho_{\rm out}$ after a series of teleportation procedures, and $\rho_{\rm out}$ can be expressed as $\rho_{\rm out}={\rm tr}_{a,A}\left[U_{t}|\psi\rangle_{a}\langle\psi|\otimes\Lambda_{AB}(\rho_{AB})U_{t}^{\dag}\right]$. In the expression, ${\rm tr}_{a,A}$ is the partial trace over the qubits $a$ and $A$, and both of them are in Alice's side. $U_{t}=\mathcal{C}_{aB}^{Z}\mathcal{C}_{AB}^{X}\mathcal{H}_{a}\mathcal{C}_{aA}^{X}$ is the unitary operator \cite{Man}, and $\mathcal{C}_{ij}^{k}(ij=aB,AB,aA;\ k=Z,X)$ denotes the controlled-$k$ operation with $i$ being the controlled qubit and $j$ being the target qubit. The Hadamard operation on qubit $a$ is denoted as $\mathcal{H}_{a}$. The teleportation fidelity $F(\alpha,\beta)$ is the overlap between the unknown input state $|\psi\rangle$ and the teleported state $\rho_{\rm out}$
\begin{equation}
F(\alpha,\beta)=\langle\psi|\rho_{\rm out}|\psi\rangle.
\end{equation}
In order to get rid of $\alpha$ and $\beta$ on the teleportation fidelity, the average teleportation fidelity is given
\begin{equation}
F=\frac{1}{4\pi}\int_{0}^{2\pi}{\rm d}\beta\int_{0}^{\pi}\sin{\alpha}F(\alpha,\beta){\rm d}\alpha,
\end{equation}
where $4\pi$ is the solid angle. Henceforth, it means the average teleportation fidelity as we refer to the teleportation fidelity. After straightforward calculation, the teleportation fidelity reads
\begin{equation}
F(\Lambda_{AB}(\rho_{AB}))=\frac{1}{6}[4+(-2+p)p+2(-1+p)^2\cos{(2\theta)}].
\end{equation}
However, the expression of relative entropy of coherence $C_{\rm re}({\rm tr}_{B}[\Lambda_{AB}(\rho_{AB})])$ is analytically messy. Alternatively, we plot the evolution of $h(F)+C_{\rm re}(\rho_{A})$, $h(F)$ and $C_{\rm re}(\rho_{A})$ as functions of the state parameter $\theta$ and the noise parameter $p$ in Fig.~\ref{fig3}. In this paragraph, $h\left(\frac{1+\sqrt{1-[3F(\Lambda_{AB}(\rho_{AB}))-2]^2}}{2}\right)$ and $C_{\rm re}({\rm Tr}_{B}[\Lambda_{AB}(\rho_{AB})])$ are denoted by $h(F)$ and $C_{re}(\rho_{A})$ for the sake of simplicity in the case that it does not cause confusion. From the figure, it is found that $h(F)$ and $C_{\rm re}(\rho_{A})$ compensate each other. For a fixed value of $p$, the relative entropy of coherence $C_{\rm re}(\rho_{A})$ increases when $h(F)$ decreases with the increasing of $\theta$, or vice verse. These results can be observed much more clearly from Fig.~\ref{fig4}, where the evolutions of $h(F)+C_{\rm re}(\rho_{A})$, $h(F)$ and $C_{\rm re}(\rho_{A})$ versus $\theta$ for a fixed value of $p=0$ are plotted. The underlying physical mechanism for these results is that the resource state changes from the maximally entangled state $|\Phi\rangle_{AB}$ to the product state $\frac{1}{\sqrt{2}}(|0\rangle+|1\rangle)_{A}\otimes\frac{1}{\sqrt{2}}(|0\rangle+|1\rangle)_{B}$ when $\theta$ ranges from $0$ to $\pi/2$. The maximally entangled state can be used for teleportation with the fidelity getting the maximum value $1$, however, the relative entropy of coherence of the subsystem $A$
equals to zero. On the contrary, the product state cannot be used for teleportation while $C_{\rm re}(\rho_{A})=1$.

As proved in Ref.~\citen{Xi}, the relative entropy of coherence is unitary invariant by using the different bases, the results given in Eqs.~(\ref{eq3}) and~(\ref{eq13}) hold for all local bases.

From the results given in Eqs.~(\ref{eq3}) and ~(\ref{eq13}), it is found that there is trade-off between local quantum coherence and the optimal dense coding capacity or the teleportation fidelity. In general, the relation among coherence, discord and entanglement has been given by use of quantum relative entropy, where quantum coherence is found to be a more ubiquitous manifestation of quantum correlations \cite{Yao}. For two-qubit states with maximally mixed marginals, the pairwise correlations between local observables are complementary to the coherence of the product bases they define~\cite{Giorda}. Furthermore, the results in Ref.~\citen{Yao,Giorda} also indicate that the existence of correlations, particularly entanglement, together with the purity of the global state, implies that the reduced states are highly mixed, and thus have low coherence in any basis. Combing with the fact that dense coding and teleportation rely on quantum correlations, especially entanglement, our complementarity relations between local quantum coherence and dense coding capacity or teleportation fidelity can be easily understood. Therefore, our results in the present paper are harmonious with those given in Ref.~\citen{Yao,Giorda}.

\section*{Discussion}
{\color{red}{In this paper, we relate the relative entropy of coherence to quantum dense coding and teleportation. Firstly, we establish a complementarity relation between the optimal dense coding capacity of a bipartite system and local quantum coherence. The inequality indicates that smaller local quantum coherence will bring about the greater capacity of optimal dense coding. It is also found that the relation can be applied to the case of dense coding by using unital memoryless noisy quantum channels. Secondly, an inequality in the form of complementarity relation between teleportation fidelity for a two-qubit system and local quantum coherence of its subsystem is given. From the inequality, it is found that the greater the teleportation fidelity is, the smaller local quantum coherence will be. Our results in this paper give a clear quantitative analysis between quantum coherence and some specific quantum information protocols.}}

{\color{red}{In the subsection of relating quantum coherence to optimal dense coding, it is found that the result given in Theorem 1 can also be extended to the case of dense coding by using unital memoryless noise quantum channels. In general, our results given in Eqs.~(\ref{eq3}) and ~(\ref{eq13}) can be generalized to general noisy maps. A noisy map can be described by a completely positive trace preserving linear map $\Lambda(\rho)=\sum_{i}K_{i}\rho K_{i}^{\dag}$ with the Kraus operators $K_{i}$ satisfying $\sum_{i}K_{i}^{\dag}K_{i}=I$. If $\rho_{AB}$, $\rho_{A}$ and $\rho_{B}$ are respectively substituted by $\Lambda_{AB}(\rho_{AB})$, ${\rm tr}_{B}(\Lambda_{AB}(\rho_{AB}))$ and ${\rm tr}_{A}(\Lambda_{AB}(\rho_{AB}))$, the results given in Eqs.~(\ref{eq3}) and ~(\ref{eq13}) are still tenable. Actually, in the subsection of relating quantum coherence to teleportation, we have considered the distribution of two-qubit state through $2$-dimensional depolarizing channels, and found that the Eq.~(\ref{eq13}) is still satisfied.}}

%\bibliography{sample}

\begin{thebibliography}{99}

\bibitem{Nielsen} Nielsen, M. A. \& Chuang, I. L. Quantum Computation and Quantum Information (Cambridge University Press, 2000).

\bibitem{Demkowicz} Demkowicz-Dobrza\'{n}ski, R. \& Maccone, L. Using entanglement against noise in quantum metrology. Phys. Rev. Lett. \textbf{113}, 250801 (2014).

\bibitem{Giovannetti1} Giovannetti, V., Lloyd, S. \& Maccone, L. Advances in quantum metrology. Nat. Photonics \textbf{5}, 222-229 (2011).

\bibitem{Aberg} {\AA}berg, J. Catalytic coherence. Phys. Rev. Lett. \textbf{113}, 150402 (2014).

\bibitem{Narasimhachar} Narasimhachar, V. \& Gour, G. Low-temperature thermodynamics with quantum coherence. Nat. Commun. \textbf{6}, 7689 (2015).

\bibitem{Cwiklinski} \'{C}wikli\'{n}ski, P., Studzi\'{n}ski, M., Horodecki, M. \& Oppenheim, J.  Limitations on the evolution of quantum coherences: Towards fully quantum second laws of thermodynamics. Phys. Rev. Lett. \textbf{115}, 210403 (2015).

\bibitem{Lostaglio} Lostaglio, M., Jennings, D. \& Rudolph, T. Description of quantum coherence in thermodynamic processes requires constraints beyond free energy. Nat. Commun. \textbf{6}, 6383 (2015).

\bibitem{Lostaglio1} Lostaglio, M., Korzekwa, K., Jennings, D. \& Rudolph, T. Quantum coherence, time-translation symmetry, and thermodynamics. Phys. Rev. X \textbf{5}, 021001 (2015).

\bibitem{Plenio} Plenio, M. B. \& Huelga, S. F. Dephasing-assisted transport: quantum networks and biomolecules. New J. Phys. \textbf{10}, 113019 (2008).

\bibitem{Lloyd} Lloyd, S. Quantum coherence in biological systems. J. Phys: Conf. Ser. \textbf{302}, 012037 (2011).

\bibitem{Li} Li, C. M. et al. Witnessing quantum coherence: from solid-state to biological systems. Sci. Rep. \textbf{2}, 885 (2012).

\bibitem{Huelga} Huelga, S. F. \& Plenio, M. B. Vibrations, quanta and biology. Contemp. Phys. \textbf{54}, 181-207 (2013).

\bibitem{Baumgratz} Baumgratz, T., Cramer, M. \& Plenio, M. B. Quantifying coherence. Phys. Rev. Lett. \textbf{113}, 140401 (2014).

\bibitem{Streltsov} Streltsov, A. et al. Measuring quantum coherence with entanglement. Phys. Rev. Lett. \textbf{115}, 020403 (2015).

\bibitem{Winter} Winter, A. \& Yang, D. Operational resource theory of coherence. Phys. Rev. Lett. \textbf{116}, 120404 (2016).

\bibitem{Chitambar} Chitambar, E. et al. Assisted distillation of quantum coherence. Phys. Rev. Lett. \textbf{116}, 070402 (2016).

\bibitem{Yuan} Yuan, X., Zhou, H., Cao, Z. \& Ma, X. Intrinsic randomness as a measure of quantum coherence. Phys. Rev. A \textbf{92}, 022124 (2015).

\bibitem{Du} Du, S., Bai, Z. \& Qi, X. Coherence measures and optimal conversion for coherent states.  Quantum Inf. Comput.\textbf{15}, 1355-1364 (2015).

\bibitem{Yao} Yao, Y., Xiao, X., Ge, L. \& Sun, C. P. Quantum coherence in multipartite systems. Phys. Rev. A \textbf{92}, 022112 (2015).

\bibitem{Xi} Xi, Z., Li, Y. \& Fan, H. Quantum coherence and correlations in quantum system. Sci. Rep. \textbf{5}, 10922 (2015).

\bibitem{Cheng} Cheng, S. \& Hall, M. J. W. Complementarity relations for quantum coherence. Phys. Rev. A \textbf{92}, 042101 (2015).

\bibitem{Streltsov3} Streltsov, A. Genuine quantum coherence. J. Phys. A: Math. Theor. \textbf{50}, 045301 (2017).

\bibitem{deVicente} de Vicente, J. I. \& Streltsov, A. The power of the resource theory of genuine quantum coherence. Preprint at https://arxiv.org/abs/1604.08031.

\bibitem{Streltsov1} Streltsov, A. et al. Entanglement and coherence in quantum state merging. Phys. Rev. Lett. \textbf{116}, 240405 (2016).

\bibitem{Streltsov2} Streltsov, A., Rana, S., Bera, M. N. \& Lewenstein, M. Hierarchies of incoherent quantum operations. Preprint at https://arxiv.org/abs/1509.07456.

\bibitem{Marvian} Marvian, I., Spekkens, R. W. \& Zanardi, P. Quantum speed limits, coherence, and asymmetry. Phys. Rev. A \textbf{93}, 052331 (2016).

\bibitem{Bennett} Bennett, C. H. \& Wiesner, S. J. Communication via one- and two-particle operators on Einstein-Podolsky-Rosen states. Phys. Rev. Lett. \textbf{69}, 2881-2884 (1992).

\bibitem{Bennett1} Bennett, C. H. et al. Teleporting an unknown quantum state via dual classical and Einstein-Podolsky-Rosen channels. Phys. Rev. Lett. \textbf{70}, 1895-1899 (1993).

\bibitem{Addis} Addis, C., Brebner, G., Haikka, P. \& Maniscalco S. Coherence trapping and information backflow in dephasing qubits. Phys. Rev. A \textbf{89}, 024101 (2014).

\bibitem{Bromley} Bromley, T. R., Cianciaruso, M. \& Adesso, G. Frozen quantum coherence. Phys. Rev. Lett. \textbf{114}, 210401 (2015).

\bibitem{Yu} Yu, X.-D., Zhang, D.-J., Liu, C. L. \& Tong, D. M. Measure-independent freezing of quantum coherence. Phys. Rev. A \textbf{93}, 060303 (2016).

\bibitem{Bhattacharya} Bhattacharya, S., Banerjee, S. \& Pati, A. K. Effect of non-Markovianity on the dynamics of coherence, concurrence and Fisher information. Preprint at https://arxiv.org/abs/1601.04742.

\bibitem{Shadman} Shadman, Z., Kampermann, H., Macchiavello, C. \& Bru{\ss}, D. Optimal super densce coding over noisy quantum channels. New J. Phys. \textbf{12}, 070342 (2010).

\bibitem{Shadman1} Shadman, Z., Kampermann, H., Bru{\ss}, D. \& Macchiavello, C. Optimal superdense coding over memory channels. Phys. Rev. A \textbf{84}, 042309 (2011).

\bibitem{Shadman2} Shadman, Z., Kampermann, H., Bru{\ss}, D. \& Macchiavello, C. Distributed superdense coding over noisy channels. Phys. Rev. A \textbf{85}, 052306 (2012).

\bibitem{Das} Das, T., Prabhu, R., Sen(De), A. \& Sen, U. Multipartite dense coding versus quantum correlation: Noise inverts relative capability of information transfer. Phys. Rev. A \textbf{90}, 022319 (2014).

\bibitem{Das1} Das, T., Prabhu, R., Sen(De), A. \& Sen, U. Distributed quantum dense coding with two receivers in noisy environments. Phys. Rev. A \textbf{92}, 052330 (2015).
    
\bibitem{Wangxin} Wang, X. et al. Relating quantum discord with the quantum dense coding capacity. J. Exp. Theor. Phys. \textbf{120}, 9-14, (2015).

\bibitem{Liu} Liu, B. H. et al. Efficient superdense coding in the presence of non-Markovian noise. EPL \textbf{114}, 10005 (2016).

\bibitem{Bowen} Bowen, G. \& Bose, S. Teleportation as a depolarizing quantum channel, relative entropy, and classical capacity. Phys. Rev. Lett. \textbf{87}, 267901 (2001).

\bibitem{Albeverio} Albeverio, S., Fei, S.-M. \& Yang, W.-L. Optimal teleportation based on bell measurements. Phys. Rev. A \textbf{66}, 012301 (2002).

\bibitem{Oh} Oh, S., Lee, S. \& Lee, H.-W. Fidelity of quantum teleportation through noisy channels. Phys. Rev. A \textbf{66}, 022316 (2002).

\bibitem{Taketani} Taketani, B. G., de Melo, F. \& de Matos Filho, R. L. Optimal teleportation with a noisy source. Phys. Rev. A \textbf{85}, 020301(R) (2012).

\bibitem{Knoll} Knoll, L. T., Schmiegelow, C. T. \& Larotonda, M. A. Noisy quantum teleportation: An experimental study on the influence of local environments. Phys. Rev. A \textbf{90}, 042332 (2014).

\bibitem{Fortes} Fortes, R. \& Rigolin, G. Fighting noise with noise in realistic quantum teleportation. Phys. Rev. A \textbf{92}, 012338 (2015).

\bibitem{Fortes1} Fortes, R. \& Rigolin, G. Probabilistic quantum teleportation in the presence of noise. Phys. Rev. A \textbf{93}, 062330 (2016).

\bibitem{Hiroshima} Hiroshima, T. Optimal dense coding with mixed state entanglement. J. Phys. A: Math. Theor. \textbf{34}, 6907-6912 (2001).
 
\bibitem{Qiu2017} Qiu, L., Wang, A. M. \& Ma, X. S. Optimal dense coding with thermal entangled states. Physica A \textbf{383}, 325-330, (2007).

\bibitem{Verstraete} Verstraete, F. \& Verschelde, H. Fidelity of mixed states of two qubits. Phys. Rev. A \textbf{66}, 022307, (2002).

\bibitem{Verstraete1} Verstraete, F., Audenaert, K., Dehaene, J. \& De Moor, B. A comparison of the entanglement measures negativity and concurrence. J. Phys. A: Math. Theor. \textbf{34}, 10327-10332 (2001).

\bibitem{Man} Man, Z.-X. \& Xia, Y.-J. Quantum teleportation in a dissipative environment. Quantum Inf. Process. \textbf{11}, 1911-1920, (2012).

\bibitem{Giorda} Giorda, P. \& Allegra, M. Two-qubit correlations revisited: average mutual information, relevant (and useful) observables and an application to remote state preparation. Preprint at https://arxiv.org/abs/1606.02197.

\end{thebibliography}

\section*{Acknowledgements}

The authors are grateful to the anonymous referees for their comments and suggestions. This work was supported by the National Natural Science Foundation of
China under Grant No. 61401465 and the Foundation Research Project (Natural Science Foundation) of Jiangsu Province under Grant No. BK20140214.

\section*{Author contributions statement}
F.P. and Z.L. initiated the research project and established the main results under the guidance of L.Q.  F.P. wrote the manuscript and all authors reviewed the manuscript.

\section*{Additional information}
\textbf{Competing financial interests:} The authors declare no competing financial interests.

\newpage

\end{document}